\documentclass[aps,prb,superscriptaddress,floatfix,showkeys]{revtex4-2}  
\usepackage{etoolbox}
\usepackage{graphics,graphicx} 
\usepackage{natbib}
\usepackage{subfig}
\usepackage{float}
\usepackage{placeins}
\usepackage{ragged2e}
\usepackage{caption}
\usepackage{xcolor}
\usepackage{amsmath}
\usepackage{amssymb}
\usepackage{tikz}
\usepackage{braket}
\usepackage{blindtext}
\usepackage{bm}
\usepackage{subfig}
\usepackage{algorithm}
\usepackage{algpseudocode}
\usepackage{hyperref}
\hypersetup{
    colorlinks=true,
    citecolor=red,
    linkcolor=blue,
    urlcolor=black
}
\usepackage{MnSymbol}
\usepackage[cal=boondox,scr=boondoxo]{mathalfa}
\usepackage{booktabs}
\begin{document}


\title{  Topologically Protected Learning from Exceptional Point Braiding: Toward Braid Programming}


\author{M.N. Jipdi}
\email[]{jmichaelnicky@yahoo.fr}
\affiliation{Department of Physics, Higher Teachers Training College, University of Bamenda, P.O.BOX: 11 Bamenda, Cameroon}
\affiliation{Unité de Recherche de Matière Condensée, d'\'Electronique et de Traitement de Signal (URMACETS), Department of Physics, Faculty of Science, University of Dschang, Cameroon, P.O.Box: 67 Dschang-Cameroon}
\affiliation{Promotion Centre of Research for Technological Advancement and Sustainable Development (PCR-TASD), P.O.Box: 55 Maroua-Cameroon}
\affiliation{Quantum Materials and Computing Group (QMaCG), P.O.Box: 70 Bambili, North West Region, Cameroon}

\author{C. Avomo Mba}
\affiliation{Laboratoire d'Optique, Laser et Applications; Universit\'e des Sciences et Techniques de Masuku, B. P. 943, Franceville, Gabon.}
\affiliation{ D\'epartement de Physique; Facult\'e des Sciences; Universit\'e des Sciences et Techniques de Masuku, B. P. 943, Franceville, Gabon.}

\author{A. B. Moubissi}
\affiliation{Laboratoire d'Optique, Laser et Applications; Universit\'e des Sciences et Techniques de Masuku, B. P. 943, Franceville, Gabon.}
\affiliation{ D\'epartement de Physique; Facult\'e des Sciences; Universit\'e des Sciences et Techniques de Masuku, B. P. 943, Franceville, Gabon.}

\author{L.C. Fai}
\affiliation{Unité de Recherche de Matière Condensée, d'\'Electronique et de Traitement de Signal (URMACETS), Department of Physics, Faculty of Science, University of Dschang, Cameroon, P.O.Box: 67 Dschang-Cameroon}
\affiliation{Quantum Materials and Computing Group (QMaCG), P.O.Box: 70 Bambili, North West Region, Cameroon}

\author{M.E. Ateuafack}
\affiliation{Department of Electrical and Electronics Engineering, College Of Technology, University of Buea, P.O. BOX: 63 Buea, Cameroon}
\affiliation{Unité de Recherche de Matière Condensée, d'\'Electronique et de Traitement de Signal (URMACETS), Department of Physics, Faculty of Science, University of Dschang, Cameroon, P.O.Box: 67 Dschang-Cameroon}
\affiliation{Promotion Centre of Research for Technological Advancement and Sustainable Development (PCR-TASD), P.O.Box: 55 Maroua-Cameroon}
\affiliation{Quantum Materials and Computing Group (QMaCG), P.O.Box: 70 Bambili, North West Region, Cameroon}


\date{\today}

\begin{abstract}
We present a framework for topological learning based on exceptional point (EP) braiding in a non‑Hermitian Bogoliubov–de Gennes Hamiltonian.
A closed algebraic equation for the EP super‑surface is derived; through momentum quantisation in finite systems, it predicts the exact number and parameter positions of all EPs in real space, irrespective of system size.
The EP topology is characterised by two quantised invariants the state‑swap fidelity \(T_{\rm swap}=0,\pm1\) and the normalised Berry phase \(\Gamma_{\rm Berry}=0,\pm\pi/2\) which cannot both be zero for a topological EP.
A complete topological map shows that all EPs lie within the region \(R_{\rm Topo} = \{(\gamma, J_1, a) \;|\; \gamma^{2} < J_{1}^{2} \;\lor\; a = 0\}\).
Adiabatic encirclements confirm robust state swapping and yield a universal set of braid gates, including Pauli‑\(X\), Pauli‑\(Y\), Pauli‑\(Z\), a Hadamard‑like gate, the \(T\)‑gate, and a SWAP operation, with the special case \(a=0\) providing additional phase gates.
Building on these generators, we reformulate learning as braid programming a discrete search over the braid group that replaces gradient descent on continuous weights with combinatorial optimisation.
A proof‑of‑concept genetic search successfully discovers short braid words that reproduce the standard Hadamard gate and the \(H\cdot Z\) gate with perfect fidelity.
This paradigm offers inherent noise immunity, catastrophic‑forgetting prevention through compositional concatenation, and guaranteed generalisation by mathematical construction, establishing EP braiding as a promising substrate for robust, interpretable, and topologically protected neuromorphic computation.

\end{abstract}


\maketitle


\section{Introduction}

Machine learning has become one of the most transformative approach with deep neural networks achieving remarkable performance on tasks
ranging from image recognition to natural and large language processing ~\cite{lecun2015deep, krizhevsky2012imagenet}.  At the heart
of virtually all state‑of‑the‑art models lies a conceptually simple
engine: optimisation of a loss function by gradient descent ~\cite{rosenblatt1958perceptron, rumelhart1986learning}.  Whether
explicitly computed or approximated, gradients guide the adjustment of
millions of continuous weights, and it is this continuous optimisation that
ultimately endows the network with its predictive power ~\cite {lecun2015deep, rumelhart1986learning}. 

The dominance of gradient‑based optimisation traces its origins to the
foundational work of Rosenblatt~\cite{rosenblatt1958perceptron}, who
introduced the perceptron as the first trainable neural network using a
simple error-correction learning rule.  The modern era of deep learning began
with the development of the backpropagation algorithm by Rumelhart
\textit{et al.}~\cite{rumelhart1986learning}, which enabled efficient
computation of gradients through multi-layer networks via the chain rule.
This breakthrough, building on earlier insights by Werbos~\cite{werbos1974beyond}
on automatic differentiation, established gradient descent as the central
algorithmic paradigm for neural network training.  The subsequent decades
witnessed an explosion of gradient‑based optimisation techniques:
Kingma \textit{et al.}~\cite{kingma2014adam} introduced the Adam optimiser,
which combines momentum and adaptive learning rates and has become a
standard for training deep networks; Sutskever \textit{et al.}~\cite{sutskever2013importance}
demonstrated the importance of momentum and careful initialisation for very
deep architectures; and Ioffe \textit{et al.}~\cite{ioffe2015batch} developed
batch normalisation, which dramatically accelerated training by reducing
internal covariate shift.  These innovations, coupled with the availability
of large‑scale datasets~\cite{deng2009imagenet} and GPU computing, enabled
the deep learning revolution that has transformed artificial intelligence.

 Yet the very mechanism that makes deep learning so successful also sows the seeds of its
most stubborn limitations. Hochreiter~\cite{hochreiter1991untersuchungen}
identified the vanishing gradient problem, which restricts the depth of
trainable networks.  Goodfellow \textit{et al.}~\cite{goodfellow2014explaining}
demonstrated that adversarial perturbations imperceptible to humans could
arbitrarily manipulate network outputs, revealing the fragility of
gradient-learned decision boundaries.  Kirkpatrick
\textit{et al.}~\cite{kirkpatrick2017overcoming} documented catastrophic
forgetting, whereby sequential learning of new tasks overwrites previously
acquired knowledge.  Zhang \textit{et al.}~\cite{zhang2017understanding}
showed that deep networks can perfectly memorise random labels, questioning
whether gradient descent truly discovers meaningful representations or
merely interpolates.  These challenges share a common origin:
gradient‑based learning encodes information metrically, in the
continuous magnitudes of weight parameters, providing no mechanism for
discrete, noise‑immune information storage.

Moreover, the extension of neural network concepts to quantum computing
introduced additional gradient-related difficulties.  Kak~\cite{kak1995quantum}
proposed early concepts of quantum neural computation, while Farhi
\textit{et al.}~\cite{farhi2018classification} formalised the quantum neural
network paradigm using parameterised quantum circuits.  Schuld
\textit{et al.}~\cite{schuld2019evaluating} analysed the gradient landscape
of quantum circuits, discovering the \textit{barren plateau} phenomenon gradients
vanish exponentially with the number of qubits over large portions of
parameter space and McClean \textit{et al.}~\cite{mcclean2018barren} showed
that this arises from the concentration of measure in high‑dimensional
Hilbert spaces, fundamentally limiting the scalability of gradient‑based
quantum machine learning.  Although alternative paradigms such as quantum
algorithms for linear systems~\cite{harrow2009quantum} and quantum kernel
methods~\cite{havlivcek2019supervised} have been proposed to circumvent these
issues, the fundamental tension remains: both classical and known quantum
neural networks rely on continuous parameter optimisation, inheriting their
associated fragility and scaling limitations.

Recently, the application of Neural Quantum States (NQS) to non-Hermitian (NH) systems has begun to emerge as a promising research direction ~\cite{carleo2017solving, wah2025manybody, wah2026bridging, wah2026autoregressive}. Although gradient-based optimisation currently constitutes the predominant approach for investigating this class of systems, the intrinsic structure of non-Hermitian physics also paves the way for a fundamentally different methodological paradigm, rooted in the physics of exceptional points.  Exceptional points (EPs) are spectral degeneracies unique to
non‑Hermitian systems where both eigenvalues and eigenvectors
coalesce; their defining characteristic is the defective nature of the
Hamiltonian, which cannot be diagonalised and instead assumes a Jordan block
structure.  The concept originated in the mathematical work of
Kato~\cite{kato1966perturbation} on perturbation theory for linear operators,
where defective eigenvalues were identified as points at which the
Hamiltonian cannot be diagonalised.  Berry~\cite{berry2004physics} provided
the first comprehensive physical treatment, demonstrating that encircling an
EP in parameter space induces eigenvalue permutation and geometric phase
accumulation phenomena with no Hermitian analogue.  Experimental
verification followed rapidly: Dembowski \textit{et al.}~\cite{dembowski2001experimental}
observed EP encirclement in a microwave cavity, confirming the predicted
eigenvalue swap; Doppler \textit{et al.}~\cite{doppler2016dynamically}
demonstrated dynamical encirclement of EPs in coupled optical waveguides,
revealing asymmetric mode switching that depends on the encirclement
direction; and Xu \textit{et al.}~\cite{xu2016topological} established the
connection between EPs and topological phases, showing that EPs can act as
topological defects in parameter space.  The fusion of EP physics with band
topology subsequently gave birth to the field of non‑Hermitian topological
phases, in which it has been established that EPs are not merely isolated
curiosities but fundamental organising principles for non‑Hermitian
topology~\cite{shen2018topological,bergholtz2021exceptional,wah2026mbme, ATEUAFACK2025130216, ding2022non}.

The potential of EPs for information processing was recognised in several
pioneering works.  Zhong \textit{et al.}~\cite{zhong2021braiding} demonstrated
theoretically that EPs can be braided in parameter space, generating
non‑Abelian holonomies analogous to those of anyons in topological quantum
computation~\cite{nayak2008non}.  Patil \textit{et al.}~\cite{patil2022measuring}
provided the first experimental measurement of the non‑Abelian Berry phase
associated with EP encirclement in a superconducting circuit, and
Zhang \textit{et al.}~\cite{zhang2023non} demonstrated non‑Abelian braiding
of light in coupled optical waveguides, opening the door to photonic
topological computation.  These developments suggest a tantalising
possibility: EP braiding could serve as a computational primitive for
learning.  The non‑Abelian holonomy generated by EP encirclement possesses
four essential properties: it is discrete (depending only on the
homotopy class of the trajectory), robust (invariant under
perturbations that do not cross an EP), non‑commutative (different
braid sequences yield distinct outcomes, enabling universal computation),
and geometric (encoded in the global geometry of parameter space
rather than in local parameter values) ~\cite{Pap2018Nonabelian, Tang2021Experimental, shan2024non, Park2025Singularity}.  These properties directly address
the fundamental limitations of gradient‑based learning: discreteness confers
noise immunity, robustness prevents catastrophic forgetting, and geometric
encoding guarantees generalisation by construction.

The concept of a topological neural network in which computation and
learning are encoded in topological invariants rather than continuous
weights has recently begun to emerge.  Zhang \textit{et al.}~\cite{zhang2021topological}
proposed using topological invariants of data manifolds for robust
classification, while Roychowdhury \textit{et al.}~\cite{roychowdhury2023topological}
explored topological protection mechanisms for neuromorphic computing.  The
Schrieffer‑Wolff transformation, developed by Schrieffer and
Wolff~\cite{schrieffer1966relation} and extended by Bravyi
\textit{et al.}~\cite{bravyi2011schrieffer}, provides the theoretical tools
for deriving effective Hamiltonians near EPs.  However, a unified framework
that integrates EP braiding as a computational primitive for learning with
explicit Hamiltonian construction, braid gate definition, learning algorithm
design, and numerical verification protocols has remained absent from
the literature.

The present work fills this gap.  While the individual components of
exceptional point physics, non‑Abelian holonomy, topological computation,
and neuromorphic engineering have each been extensively developed, their
synthesis into a complete topological learning framework has not been
achieved.  Specifically, no prior work has (i)~defined explicit braid gates
for a concrete non‑Hermitian Hamiltonian suitable for physical
implementation, (ii)~formulated learning as braid programming with search
algorithms operating over the braid group, (iii)~demonstrated topological
selectivity the coexistence of topological and trivial computational modes
within the same physical system, (iv)~derived a closed supersurface equation
that carries all EPs of the system and, through momentum quantisation,
predicts their exact number and positions in real space, (v)~provided a
complete topological phase map identifying the region where braid‑based
learning is operational, or (vi)~established protocols for the numerical verification of Berry phases, eigenvector swaps, and
non‑Abelian statistics.  We address all of these points by presenting a
complete theoretical and computational framework for topological learning via
EP braiding.

The remainder of this paper is organised as follows.
Sec~2 presents the exact diagonalisation strategy for the non‑Hermitian
Hamiltonian using the BdG transformation and its block structure.
Sec~3 derives the EP condition and the band‑topology criterion, and
introduces the topological invariants that characterise EPs.
Sec~4 develops braiding as a computational primitive: we define the
elementary braid gates, demonstrate their non‑Abelian statistics, and
construct a universal gate set that includes Pauli‑$X$, Pauli‑$Y$,
Pauli‑$Z$, a Hadamard‑like gate, the $T$‑gate, and a SWAP operation.
Sec~5 discusses the EP topological selectivity, its connection to Riemann surfaces and its implications for
hybrid computing architectures.
Sec~6 summarises the results and outlines future directions.

\section{Exact Block-Diagonalisation via Bogoliubov-de Gennes Transformation }

Non-Hermitian Hamiltonians have revolutionised our understanding of open quantum systems, giving rise to phenomena without Hermitian counterparts: exceptional points (EPs) , skin effects, and non-Hermitian topological phases  ~\cite{banerjee2023nonhermitian, bergholtz2021exceptional, wah2026mbme}. Simultaneously, neural networks with non-Hermitian weight matrices have been shown to exhibit enhanced expressivity and faster learning ~\cite{socci2024nonnormal, arjovsky2016unitary}. Bridging these domains, we study a spin chain whose complex transverse field simulates the effective dynamics of a neural network. Our Hamiltonian consist of exchange interactions (Isotropic exchange $J_1$ and anisotropic exchange $J_2$) which favour spin alignment in the plane. The anisotropic exchange term $J_2$ introduces frustration in the ordering and simulate a chirality in the chain.
Complex transverse field  $g_j = a + i \gamma e^{i\pi j}$ is applied on the longitudinal direction and set to be site dependent. The real part $a$ is a uniform field pointing along $z$. The imaginary part $\gamma e^{i\pi j}$  breaks Hermiticity, leading to non-conservation of probability and the possibility of EPs.

Exact diagonalisation of this many-body Hamiltonian typically requires handling a $2^N \times 2^N$ matrix, which is prohibitive for large $N$. However, if the Hamiltonian can be mapped to a quadratic form in fermion operators, the problem reduces to diagonalising a single-particle matrix of size $O(N)$. Our strategy consists of Jordan-Wigner transformation that convert each network node (spins) into fermions, yielding a Hamiltonian bilinear in creation and annihilation operators; the Fourier transform, used to exploit translation invariance and travel to momentum space; the Bogoliubov-de Gennes diagonalisation help to obtain the quasiparticle spectrum.
Each step of this methodology is exact, thus the final result is predicted to be valid as a brute-force diagonalisation but exponentially more efficient.

We therefore consider a neural network where node are a one-dimensional spin-1/2 chain with XYZ-type interactions and a complex magnetic field.  The Hamiltonian on a 1D Neural network  chain with \(N\) sites is written as:

\begin{equation}
H = \sum_{j=1}^{N-1} \bigg[ J_1 \big( S_j^x S_{j+1}^x + S_j^y S_{j+1}^y \big) + J_2 \big( S_j^x S_{j+1}^y + S_j^y S_{j+1}^x \big) \bigg] + \sum_{j=1}^{N} (a + i \gamma_j) S_j^z
\end{equation}

Here: \(J_1, J_2, a\) are real constants, \(\gamma e^{i\pi j}\) are real site-dependent coefficients, \(S_j^\alpha = \frac{1}{2} \sigma_j^\alpha\) are spin-1/2 operators. The term \((a + i \gamma e^{i\pi j}) S_j^z\) introduces a non-Hermitian complex field. We aim first to map this Hamiltonian to fermionic creation and annihilation operators \(c_j^\dagger, c_j\) using the exact Jordan–Wigner transformation. For a one-dimensional chain, the Jordan–Wigner transformation is defined as:

\begin{align}
S_j^+ &= S_j^x + i S_j^y = \left( \prod_{k=1}^{j-1} \sigma_k^z \right) c_j^\dagger, \label{eq:JW_plus} \\
S_j^- &= S_j^x - i S_j^y = \left( \prod_{k=1}^{j-1} \sigma_k^z \right) c_j, \label{eq:JW_minus} \\
S_j^z &= c_j^\dagger c_j - \frac{1}{2}. \label{eq:JW_z}
\end{align}

The operators \(c_j, c_j^\dagger\) satisfy canonical fermionic anticommutation relations:
\begin{equation}
\{c_i, c_j^\dagger\} = \delta_{ij}, \qquad \{c_i, c_j\} = 0, \qquad \{c_i^\dagger, c_j^\dagger\} = 0.
\end{equation}

The Pauli matrices \(\sigma_k^z = 2 S_k^z = 2c_k^\dagger c_k - 1\) satisfy \((\sigma_k^z)^2 = 1\) and commute for different sites.
For any \(j\), we have:
\begin{align}
\sigma_j^z c_j^\dagger &= c_j^\dagger, \label{eq:sigma_cdag} \\
\sigma_j^z c_j &= -c_j, \label{eq:sigma_c} \\
\prod_{k=1}^{j-1} \sigma_k^z \prod_{k=1}^{j-1} \sigma_k^z &= I. \label{eq:string_square}
\end{align}

Making use of these transformations, the Hamiltonian of the system is rewritten in terms of Fermionic operators

\begin{equation}
H_{\text{fermion}} = \sum_{j=1}^{N} \left[ \frac{J_1}{2} \left( c_j^\dagger c_{j+1} + c_{j+1}^\dagger c_j \right) + \frac{J_2}{2i} \left( c_j^\dagger c_{j+1}^\dagger - c_{j+1}c_j  \right) \right] + \sum_{j=1}^{N} (a + i (-1)^j \gamma) c_j^\dagger c_j
\end{equation}

The omitted term $- \frac{1}{2} \sum_{j=1}^{L} (a + i  (-1)^j \gamma)$ is a constant energy shift that plays a crucial role in determining the ground state energy. In the fermionic framework, the first term  represent the hopping, the second term is the pairing term coupling particle and hole in momentum space, and the diagonal term is non-Hermitian representing gain/loss in the fermionic picture. The resulting Hamiltonian describes spinless fermions with nearest-neighbour hopping, \(p\)-wave pairing, and a complex chemical potential.

Because $(-1)^j$ alternates, we introduce two sublattices descriptions; we then split the network in even and odd sites, therefore,

\[\begin{array}{l}
{H_{{\rm{fermion}}}} = \sum\limits_{m = 1}^M {\frac{{{J_1}}}{2}\left( {c_{2m}^\dag {c_{2m + 1}} + c_{2m + 1}^\dag {c_{2m}} + c_{2m - 1}^\dag {c_{2m}} + c_{2m}^\dag {c_{2m - 1}}} \right)}  + \sum\limits_{m= 1}^M {\left( {a + i\gamma } \right)} c_{2m}^\dag {c_{2m}}\\
\,\,\,\,\,\,\,\,\,\,\,\,\,\,\,\, + \sum\limits_{m = 1}^M {\frac{{{J_2}}}{{2i}}\left( {c_{2m}^\dag c_{2m + 1}^\dag  - {c_{2m + 1}}{c_{2m}} + c_{2m - 1}^\dag c_{2m}^\dag  - {c_{2m}}{c_{2m - 1}}} \right) + \sum\limits_{m = 1}^M {\left( {a - i\gamma } \right)} c_{2m - 1}^\dag {c_{2m - 1}}}
\end{array}\]

now we set

\[
A_m = c_{2m}, \qquad B_m = c_{2m-1},\qquad m=1,\dots,M,\; N=2M.
\]
For periodic boundary conditions we take the Fourier transform on the $m$-lattice :
\[
A_m = \frac{1}{\sqrt{M}}\sum_{k} e^{i2km} A_k,\qquad
B_m = \frac{1}{\sqrt{M}}\sum_{k} e^{ik (2m-1) } B_k,
\]
with $k\in]-\pi,\pi]$ . The Hamiltonian becomes a sum over $k$ of a $4\times4$ Bogoliubov-de Gennes matrix acting on the Nambu spinor
\[
\Psi_k = \bigl( A_k,\; B_k,\; A_{-k}^\dagger,\; B_{-k}^\dagger \bigr)^T.
\]
After some  algebra,  we obtain
\[
H_{\text{BdG}}(k) =
\begin{pmatrix}
(a+i\gamma) & J_1\cos k & 0 & J_2 \sin k \\
J_1\cos k & (a-i\gamma) & J_2 \sin k & 0 \\
0 & J_2 \sin k & -(a+i\gamma) & -J_1\cos k \\
J_2 \sin k & 0 & -J_1\cos k & -(a-i\gamma)
\end{pmatrix}.
\]

The eigen values of the Hamiltonian are given by

\begin{equation}
{E_k} =  \pm {\rm{ }}\sqrt {J_1^2{{\cos }^2}\left( k \right) + {a^2} - {\gamma ^2} + J_2^2{{\sin }^2}\left( k \right) \pm {\rm{ }}\sqrt {J_1^2{a^2}{{\cos }^2}\left( k \right) - {\gamma ^2}J_2^2{{\sin }^2}\left( k \right) - {a^2}{\gamma ^2}} }
\end{equation}

where 2 branches are for particles symetry, while the other two are for hole symmetry. It is now crucial relates our Bogoliubov-de Gennes analysis to brute-force exact diagonalisation of the original many-body Hamiltonian.

It is instructive to mention that, since $H$ is quadratic, the many-body eigenstates are Slater determinants constructed from the quasiparticle operators $A_{k,\sigma}$, $B_{k,\sigma}$ that diagonalise $H$:
\begin{equation}
H = \sum_{k,\sigma} E_{k,\sigma} \,\left(  A_{k,\sigma}^\dagger A_{k,\sigma}+ B_{k,\sigma}^\dagger B_{k,\sigma}\right) + E_0,
\end{equation}
where $E_0$ is the ground state energy defined as
\begin{equation}
E_0 = -\frac{1}{2} \sum_{k,\sigma} E_{k,\sigma} - \sum_{j} g_j.
\end{equation}

or in the thermodynamic limit, the exact ground state energy per site
\begin{equation}
\frac{E_0}{N} = -\frac{1}{\pi} \int_0^{\pi} E_{k\sigma } dk - a.
\end{equation}

The many-body eigenvalues are then:
\begin{equation}
E_{\{n_{k,\sigma}\}} = E_0 + \sum_{k,\sigma} n_{k,\sigma} E_{k,\sigma},
\end{equation}
with $n_{k,\sigma} \in \{0,1\}$ (Fermi-Dirac statistics). This formula yields $2^{2N}$ distinct energies (for $N$ distinct $k$ values, two bands each), which is exactly the dimension of the fermionic Fock space. Therefore, diagonalising $\mathcal{H}_k$ for each $k$ is equivalent to exact diagonalisation of the original $2^N \times 2^N$ spin Hamiltonian but exponentially faster.

\section{Exceptional Points and Topological Invariant}

\subsection{Exceptional Points}

Exceptional points are degeneracies where not only eigenvalues but also eigenvectors coalesce, making the Hamiltonian non-diagonalizable. They are unique to non-Hermitian systems and have no Hermitian counterpart.
For the $4\times4$ matrix $ H_{\text{BdG}}(k)$, both eigenvalues $E_{k,+}$ and $E_{k,-}$ are doubly degenerate (each $\pm$ band has two orthogonal states due to Kramers-like degeneracy from particle-hole symmetry). True EPs occur when these two bands touch, and the eigenvectors align. This requires:
\begin{equation}
 J_1^2{a^2}{{\cos }^2}\left( k \right) - \gamma _0^2J_2^2{{\sin }^2}\left( k \right) - {a^2}\gamma _0^2 = 0.
\end{equation}
this helps in finding $k$ via the relation

\[
\cos^2 k = \frac{\gamma_0^2 (J_2^2 + a^2)}{J_1^2 a^2 + \gamma_0^2 J_2^2}.
\]
For an exceptional point to exist in the physical Brillouin zone ($k=\frac{2 \pi n}{N} , n \in \mathbb{N}$, $k \in -[\pi..\pi] $), this expression must satisfy $0 \le \cos^2 k \le 1$. The upper bound $\cos^2 k \le 1$ gives $\gamma_0^2 a^2 \le J_1^2 a^2$, which for $a \neq 0$ reduces to $|\gamma_0| \le |J_1|$. Hence, exceptional points appear on the real $k$-axis precisely when $|\gamma_0| \le |J_1|$ , with the equalities $|\gamma_0| = |J_1|$ or $a=0$  placing the EP at $k = 0$ (or $k = \pi$). When $|\gamma_0| > |J_1|$ and $a \neq 0$,  no real $k$ satisfies coalescence, and the EPs move into the complex $k$-plane. Thus, $|\gamma_0| = |J_1|$ marks the separating limit between the Eps proficiency regime and Eps disappearance. This special point accidentally corresponds to the closing gap limit, conditioning the existence of Eps with topological phase. In the special case of where $k=0,\pi$, the  energies branch and eigen vectors becomes

\begin{equation}
{\eta _{1 \pm }} = \frac{1}{{\sqrt 2 }}{\left( { \pm \exp \left( { \mp i\theta } \right),1,0,0} \right)^T},{\eta _{2 \pm }} = \frac{1}{{\sqrt 2 }}{\left( {0,0, \pm \exp \left( { \mp i\theta } \right),1} \right)^T}
\end{equation}
\begin{equation}
E_{1k,\pm} = a\pm \sqrt{  J_1^2 -\gamma^2 } \quad E_{2k,\pm} = -a\pm \sqrt{  J_1^2 -\gamma^2 }
\end{equation}

\[\theta  = \arctan \left( {\frac{\gamma }{{\sqrt {J_1^2 - {\gamma ^2}} }}} \right)\]

where eigen vectors and eigen values coalesce at the boundary where $|\gamma_0| = |J_1|$; for the specific case $a=0$, the particles and hole energy are degenerated, giving rise to unified band, with eigenvectors independent of system parameter and no coalesce of particle and hole states, signature of Majorana-like degeneracy. Therefore, proliferation of Eps is observed in a selective domain with the limit being coincidentally that of the closed gap that correspond to $ \gamma^2 \le J_1^2$ or $a=0$

\subsection{Topological invariant: Band topology}

For non-Hermitian systems, a more appropriate invariant is the winding number of the complex energy bands; given $H_{\text{BdG}}(k)$ the Bogoliubov De Genne Hamiltonian of the system, $H_{\text{BdG}}(k)$  possesses particle-hole symmetry $\mathcal{C} H_{\text{BdG}}(k) \mathcal{C}^{-1} = -H_{\text{BdG}}(-k)^*$, placing it in symmetry class D with a $Z_2$ topological invariant.
The matrix $H_{\text{BdG}}(k)$ can be written as a $2\times2$ blocks of :
\[
h(k) = \begin{pmatrix} 2(a+i \gamma) & 2J_1\cos k \\ 2J_1\cos k & 2(a-i\gamma) \end{pmatrix},\qquad
\Delta(k) = \begin{pmatrix} 0 & J_2\sin k  \\ J_2\sin k  & 0 \end{pmatrix}.
\]
Then
\[
H_{\text{BdG}}(k) = \begin{pmatrix}  h(k) & \Delta(k) \\ \Delta(k)^\dagger & -h(k)^T  \end{pmatrix}.
\]

The \(\mathbb{Z}_2\) invariant for class D is \(\nu/2 \bmod 2\), so a non-trivial topological phase occurs when \(\nu = \pm 2\). The winding number $\upsilon$ is defined as

\begin{equation}
\upsilon=\nu/2 \bmod 2= \frac{1}{4\pi i} \int_{-\pi}^{\pi} \frac{d}{dk} \log D\left( k \right) dk.
\end{equation}

where $D\left( k \right)= det \left( h(k)+i\Delta(k) \right)$   is the determinant of the complex matrix obtained from Bogoliubov De Genne Hamiltonian

This winding number counts the number of times the complex eigenvalues encircle a base point. For our model,
\[
D(z) = a^2 + b^2 - 2 \alpha \beta- \alpha^{2}z^2 - \beta^2 z^{-2},
\]
with \(z = e^{ik}\) and
\[
\alpha = \frac{J_1 + J_2}{2}, \qquad \beta = \frac{J_1 - J_2}{2}.
\]
The winding number (topological charge) is given by
\[
\upsilon = \frac{1}{4\pi i} \oint_{|z|=1} \frac{D'(z)}{D(z)}\,dz.
\]

Evaluating the integral via the residue theorem yields
\[
\upsilon = \frac{1}{2}n - 1,
\]
where \(n\) is the number of zeros of \(F(z) = -\alpha^2 z^4 +  \left(  a^2 + \gamma^2 - 2 \alpha \beta\right)z^2 - \beta^2\) inside the unit circle. So a non-trivial topological phase occurs when \(\upsilon= \pm 1\), i.e. when \(n = 0\) or \(n = 4\).

After analysing the roots of the quadratic of $F(z)$, one finds three distinct regimes of parameters \(a,\gamma,J_1,J_2\) : Regime I   with winding number of $\upsilon= +1$  obtained for \(a^2+\gamma^2 < J_1^2 - J_2^2\),
Regime II with $\upsilon= -1$  obtained for \(J_1^2 - J_2^2 < a^2+\gamma^2 < J_1^2\) and Regime III with $\upsilon = 0$ obtained for \(a^2+\gamma^2 > J_1^2\). Regimes I and II are topological phases with opposite winding, while regime III is a non-topological or trivial phase. In the special limit \(J_1 = J_2\), the first regime disappears and one recovers the Kitaev chain result:
\[
0 < a^2+\gamma^2 < J_1^2 \quad\Longrightarrow\quad \upsilon = -1.
\]

Surprisingly, an exceptional point (EP) occurs in the momentum-space Hamiltonian when eigen vectors coalesce that correspond to a closed gap in momentum space and resume via the condition  \(\gamma^2 \le J_1^2\) or $a=0$. Remarkably, this condition together with \(a^2+\gamma^2 \le J_1^2\) (Regimes I and II) is precisely the region where the topological charge \(\upsilon\) is non-zero. Hence, a band-topology  exist on the real k-\text{-axis} where the system exhibits exceptional points.

The band-topological invariant is computed from the winding of the off-diagonal block of the Hamiltonian in the eigenbasis of the chiral symmetry operator. The condition $a^2 + \gamma^2 < J_1^2$ ensures that the winding number is odd, corresponding to the topologically non-trivial phase. At $a^2 + \gamma^2 = J_1^2$, the gap closes, and a topological phase transition occurs. A schematic diagram illustrating band topology and Ep prolifiration is provided in Fig 1. 

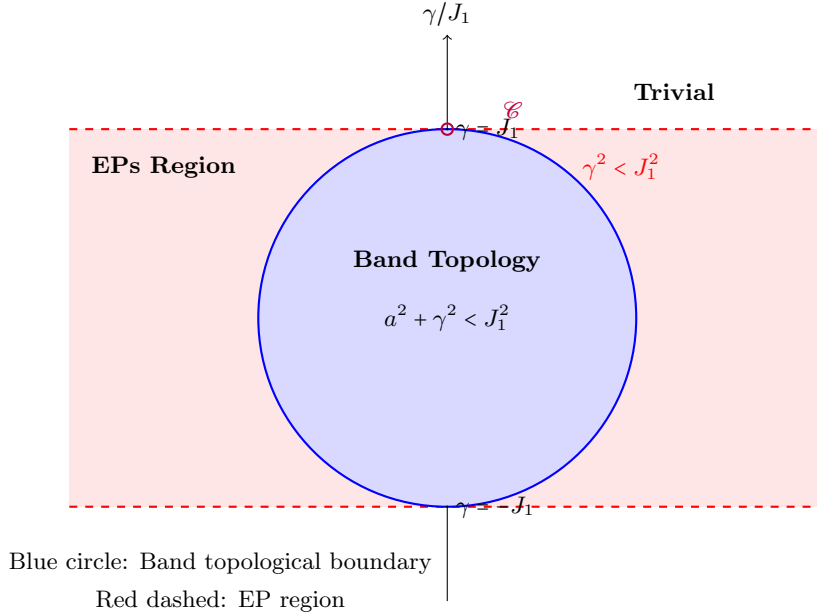
\begin{figure*}[ht]
 \centering
\begin{tikzpicture}[scale=2.5]
    \draw[->] (-1.5,0) -- (1.5,0) node[right] {$a/J_1$};
    \draw[->] (0,-1.5) -- (0,1.5) node[above] {$\gamma/J_1$};
    \fill[red!10] (-2.0,-1.0) rectangle (2.0,1.0);
    \draw[thick, red, dashed] (-2.0,-1.0) -- (2.0,-1.0);
    \draw[thick, red, dashed] (-2.0,1.0) -- (2.0,1.0);
    \fill[blue!15] (0,0) circle (1);
    \draw[thick, blue] (0,0) circle (1);
    \node at (0,0.3) {\textbf{Band Topology}};
    \node at (0,0) {$a^2+\gamma^2 < J_1^2$};
    \node[red] at (0.92,0.8) {$\gamma^2 < J_1^2$};
    \node at (1.2,1.2) {\textbf{Trivial}};
    \node at (-1.5,0.8) {\textbf{EPs Region}};
    \filldraw[black] (0,1) circle (0.1pt) node[right] {$\gamma=J_1$};
    \filldraw[black] (0,-1) circle (0.1pt) node[right] {$\gamma=-J_1$};
    \draw[thick, ->, purple] (0,1) circle (0.03);
    \node[purple] at (0.35,1.1) {$\mathcal{C}$};
    \node at (-1.2,-1.3) {\small Blue circle: Band topological boundary};
    \node at (-1.2,-1.5) {\small Red dashed: EP region };
\end{tikzpicture}
\caption{Phase diagram in the $(a, \gamma)$ plane for fixed $J_1$. The blue circle $a^2 + \gamma^2 = J_1^2$ bounds the topological region. The red dashed lines  $\gamma = \pm J_1$ mark regions where EPs occur at $k = 0, \pi$. The purple circle illustrates an encirclement path around an EP within the band topological region.}
\label{fig:phasediagram}
\end{figure*}

The observation of Fig 1 shows that the band topology erea in fully contained in the Ep Region. Moreover, the circle  \(a^2+\gamma^2 = J_1^2 - J_2^2\) separates the two topological regimes with opposite winding numbers \(\nu = +1\) and \(\nu = -1\).  This line corresponds to a reorganisation of the Band topology in the complex \(k\)-plane.

\section{Exceptional Points Braiding  for Topological Quantum Gates}
\subsection{Berry phase and Encirclement Path }
To ensure a possible braiding in the topological domain, we define an encirclement Path. An encirclement of an EP is a closed loop in parameter space that encloses the EP. For a standard adiabatic protocol, we use:
\begin{equation}
\gamma(\theta) = J_1 + R_1\cos\theta, \qquad k(\theta) = k_\text{EP} + R_2\sin\theta, \qquad \theta \in [0, 2\pi]
\end{equation}
with $R_1,R_2 \leq J_1$, while $a$ and $J_2$ are fixed. The loop lies within EPs region and  for sufficiently closed to the chosen EP so that only a single Ep is encircle. The non-Abelian holonomy resulting from adiabatic transport along a closed path $\mathcal{C}$ is:
\begin{equation}
U_{\mathcal{C}} = \mathcal{P}\exp\left(i\oint_{\mathcal{C}} A(\theta)\,d\theta\right)
\end{equation}
where $A_{mn}(\theta)$  is the Berry connection matrix in the biorthogonal basis, and $\mathcal{P}$ denotes path ordering.

To determine the Berry Phase of our  Non-Hermitian Systems, we focus on the orthogonality of eigen vectors around the EPs. The system close to an EP will have a right eigenvectors $\ket{\psi_n(\theta)}$ and left eigenvectors $\bra{\tilde{\psi}_m(\theta)}$ satisfying $\bra{\tilde{\psi}_m}\ket{\psi_n} = \delta_{mn}$, the geometric phase accumulated by state $n$ along a closed path $\mathcal{C}$ is:
\begin{equation}
\Gamma_{\rm Berry} = i\oint_{\mathcal{C}} \bra{{\tilde{\psi}}_n(\theta)}\frac{d}{d\theta}\ket{\psi_n(\theta)}\,d\theta
\end{equation}

Alternatively, we can project the resulting states obtained after evolution in an appropriate base and after projection, the accumulated phase will be obtained as a phase difference of the evolved state and the initial state. This is appropriately done using eigenvectors argument of the Holonomy Matrix (defined in the subsection bellow)

\subsection{Eigenvector Swap Detection, Eigenvalue  Permutation Tracking and Eps Topology }

In order to detect the swap of Eigen vectors, we can focus on the  Eigenvector Overlap Measure. It's quantified the degree of eigenvector exchange after one encirclement is quantified by the persistence $ O_{nn} =  \left| \bra{v_n(0)}\ket{v_n(2\pi)} \right|^2$ and the exchange $O_{nm} = \left| \bra{v_m(0)}\ket{v_n(2\pi)} \right|^2 $. For an ideal EP2 encirclement: $O_{nn} = 0$ (complete depopulation) and $O_{nm} = 1$ (complete transfer to partner).

To verify eigenvalue permutation numerically, we discretise the path: $\theta_i = 2\pi i/N$, $i = 0,\ldots, N-1$.
Diagonalize  $H_{\text{BdG}}(k(\theta_i), a, b(\theta_i), J_1, J_2)$ at each point and
sort by continuity. At $\theta_0$, we label eigenvalues and eigenvectors by their values. For $\theta_i$ ($i>0$), assign labels by mamaximising overlap. Finally, we check swap by Verifying $\lambda_+(2\pi)= \lambda_-(0)$ and $\lambda_-(2\pi) = \lambda_+(0)$ and $v_{-}(0)= v_{+}(2\pi)$, $ v_{+}(0)= v_{-}(2\pi)$

Based on this protocol, we can compute the Holonomy Matrix that is at each $\theta_i$, compute the overlap matrix between consecutive eigenbases:
    \begin{equation}
    M^{(i)}_{mn} = \ket{\tilde{v}_m(\theta_i)}\bra{v_n(\theta_{i+1})}
    \end{equation}

 The holonomy is the path-ordered product $U = \prod_{i=0}^{N-1} M^{(i)}$ ordered from right to left: $M^{(N-1)}\cdots M^{(1)}M^{(0)}$) and thus the geometric phases are the arguments of the eigenvectors of $U$. Interestingly, the holonomy matrix $U$ obtained from the adiabatic encirclement of an exceptional point along a closed path \(\mathcal{C}\) directly determines the corresponding braid gate. Its structure whether it is a permutation, a phase gate, or a coherent superposition depends on the homotopy class of \(\mathcal{C}\), the order of the exceptional point, and the geometric phases accumulated during the evolution. Thus, the holonomy matrix provides a compact representation of the unitary operation that is applied to the computational subspace when the system parameters traverse \(\mathcal{C}\).

Moreover, even if the existence of exceptional points is conditioned by
eigenvector coalescence, a defective Hamiltonian with Jordan block
structure, this alone does not guarantee that the system exhibits
an EP topology suitable for braiding operations. While eigenvector
coalescence is a necessary local condition, it is not sufficient: the EP
must also possess a non-trivial topological invariant for gates structure to hold, meaning that the
discriminant $\Delta = (\lambda_+ - \lambda_-)^2$ winds around the origin
in the complex plane when the EP is encircled. This winding defines the
Riemann surface structure of the eigenvalues, and it is non-zero only when the
encirclement path crosses the branch cut connecting the two sheets.
The system displays the characteristic eigenvalue permutation and geometric phase accumulation.

To characterise EP topology quantitatively, we focus on two complementary
observables: The Swap fidelity that quantifies the ability of the EP to exchange eigenstates upon
adiabatic encirclement. We define the swap indicator as:
\begin{equation}
\mathrm{T_{swap}} =
\begin{cases}
1 & \text{if } |\langle v_-(2\pi)|v_+(0)\rangle|^2 =
|\langle v_+(2\pi)|v_-(0)\rangle|^2 = 1,\\[4pt]
0 & \text{if } |\langle v_-(2\pi)|v_-(0)\rangle|^2 =
|\langle v_+(2\pi)|v_+(0)\rangle|^2 = 1,\\[4pt]
\text{undefined} & \text{otherwise}
\end{cases}
\end{equation}
A swap value of unity means that the state initially prepared in
$\ket{v_+(0)}$ has been completely transferred to $\ket{v_-(0)}$
(up to a geometric phase), and vice versa. This is the physical
manifestation of the eigenvalue permutation: after one complete
encirclement, the two eigenvalues have exchanged their positions
on the Riemann surface. When the swap is 0, the system doesnot exchange eigenstates but can still be topological if there is a quantised phase accumulation during encirclement. When the overlap is fractional ($0 < |\langle v_-(2\pi)|v_+(0)\rangle|^2 =
|\langle v_+(2\pi)|v_-(0)\rangle|^2< 1$), the final state is a superposition of
the two eigenstates, indicating that the encirclement path grazed the
branch cut without fully traversing it.

The second observable is the EP topological invariant, defined as the
normalised Berry phase accumulated during the encirclement:
\begin{equation}
T_{\mathrm{Berry}} = \frac{2}{\pi}\,\Gamma_{\mathrm{Berry}},
\end{equation}
where $\Gamma_{\mathrm{Berry}} $ is the non-Hermitian
Berry phase. The invariant $T_{\mathrm{Berry}}$ takes quantized values when the
encirclement is adiabatic, and the EP possesses a defined topological
charge:
\begin{equation}
T_{\mathrm{Berry}} = 0,\pm 1 \quad \Longleftrightarrow \quad
\gamma_{\mathrm{Berry}} = 0,\pm \frac{\pi}{2}.
\end{equation}
These quantised values signal that the system is in the
topological regime; the Riemann surface structure forces
the geometric phase to be a multiple of $\pi/2$, regardless of
the detailed shape or speed of the encirclement (as long as adiabaticity is maintained). The special case where Berry phase are even multiple of $\pi/2$ required a non-zero swap fidelity factor for the Ep to stand in the topological region.

Conversely, fractional values of $T_{\mathrm{Berry}}$ corresponding to a Berry phase reveal a transitional regime between trivial and topological
behaviour. In this regime, the encirclement path does not fully enclose
the branch point in the complex plane. The resulting geometric phase is not quantised because the state evolution does not project purely onto a single Riemann sheet.

Thus, the pair $(\mathrm{T_{swap}}, T_{\mathrm{Berry}})$ provides a
complete diagnostic of EP topology: ($T_\mathrm{Swap}= 1$, $T_{\mathrm{Berry}}=0,\pm 1$ ) and ($T_\mathrm{Swap}= 0$, $T_{\mathrm{Berry}}=\pm 1$ ) 
 together confirm that the system is in
a topological regime  suitable for braiding operations, while
Any deviation signals either a trivial topology or a non-adiabatic
contamination. The occurrence of each of these phases is controlled by system parameters, and a deep investigation is needed to locates EPs where braiding is effective.

\section{Results and Discussions}

In this section, we present numerical evidence for the exceptional point (EP)
topology in the non-Hermitian BdG Hamiltonian. We systematically verify
the three necessary conditions for braiding functionality: (i)~eigenvector
coalescence at the EP, (ii)~eigenstate swapping after a full adiabatic
encirclement, and (iii)~ quantisation of the geometric phase. We then construct a complete topological map identifying
all parameter regions where braiding is effective. We end the discussion by analysing braid gates and learning via braid programming.

\subsection{Confirmation of the Exceptional Point Super-Surface via Real-Space Numerical Analysis}

\begin{figure}
 \centering
\includegraphics[width=17cm,height=15cm]{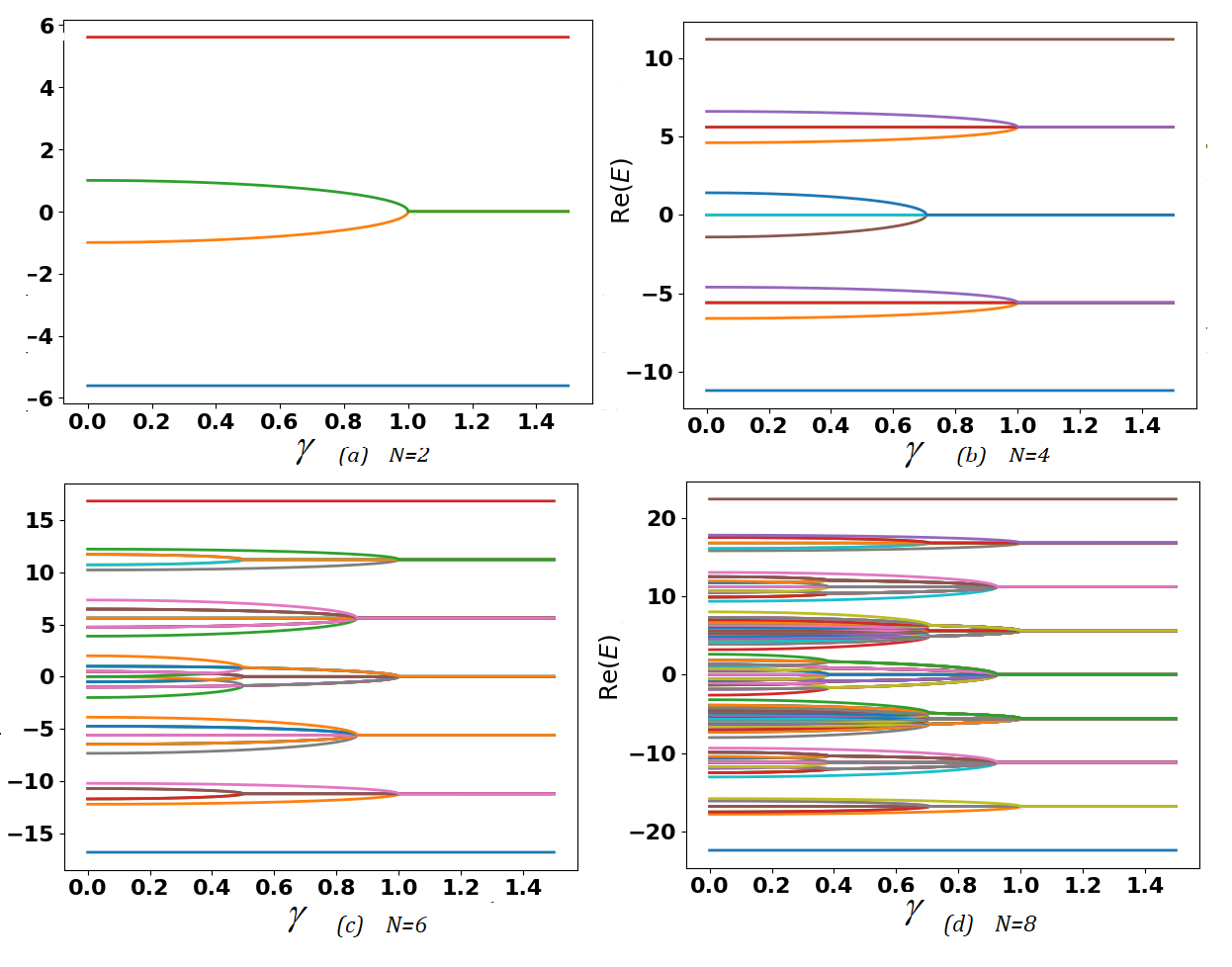}
\caption{Eigenvalue coalescence and exceptional point detection in real space for $a = 3.5J_1$, $J_2 = 0.3J_1$, $J_1 = 1$. 
(a) System with $N = 2$ spins. 
(b) System with $N = 4$ spins. 
(c) System with $N = 6$ spins. 
(d) System with $N = 8$ spins.}
\label{fig:figure1}
\end{figure}

Before analysing the topological properties of the exceptional points (EPs), it is necessary to verify their existence directly in real space. To this end, we employ exact diagonalisation of the real-space Hamiltonian for finite systems with $N = 2$, $4$, $6$, and $8$ unit cells. The coalescence of eigenvalues, which signals the presence of EPs, is presented in Fig.~\ref{fig:figure1}, where panels (a), (b), (c), and (d) correspond to $N = 2$, $4$, $6$, and $8$, respectively. The parameters are fixed at $a = 3.5J_1$, $J_2 = 0.3J_1$, with $J_1 = 1$.

To establish a direct correspondence between the $k$-space description and the real-space numerical analysis, we recall that in the momentum-space formulation, all EPs reside on a specific layer of the super-surface defined by
\begin{equation}
\cos^2 k = \frac{\gamma_0^2 (J_2^2 + a^2)}{J_1^2 a^2 + \gamma_0^2 J_2^2}.
\label{eq:ep_supersurface}  
\end{equation}
In a finite system, the momentum $k$ is quantised and takes only discrete values determined by the number of sites $N$. For $N = 2$,  EP occurs at $\left|\gamma_0\right| = \left|J_1\right|$, which corresponds to a single point $k = \pi$ in $k$-space. For $N = 4$, two EPs are found, located at $k = \pi/4$ and $k = \pi$. For $N = 6$, three EPs emerge at $k = \pi/6$, $\pi/3$, and $\pi$. For $N = 8$, four EPs are identified at $k = \pi/8$, $\pi/4$, $3\pi/8$, and $\pi$. Two central observations follow from this analysis. First, every EP obtained from exact diagonalisation lies precisely on a layer of the $k$-space super-surface given by Eq.~\eqref{eq:ep_supersurface}. Second, this super-surface captures all EPs of the system, irrespective of the system size.

This leads to a crucial result: the complete set of EPs for an arbitrary system of $N$ spin nodes is quantised along the layers of the EP super-surface. The quantisation rule emerges naturally from the Fourier transform and is given by
\begin{equation}
k = \pm \sigma \frac{\pi}{N}, \qquad \sigma \in \left[1,\frac{N}{2}\right] \cup \left[\frac{N}{2},N\right],
\label{eq:quantization}
\end{equation}

where the value $\sigma = N/2$ is excluded because it renders the system purely Hermitian, thereby preventing eigenstate coalescence. Owing to the parity of the cosine function, two distinct quantised $k$ values correspond to a single EP. Consequently, the total number of EPs in a system with $N$ spin nodes is $N/2$ for $a \neq 0$. It is important to note, however, that this counting does not account for the degeneracy of EPs in real space: several distinct real‑space configurations can map to the same $k$‑space EP, a feature that has not been included in the present enumeration but well considered by Eq~26 with $k$-values given in Eq~27.

The perfect agreement between the exact diagonalization of the real-space and the super-surface of the $k$-space is not merely a consistency check; it unveils a profound structural property of the BdG system. The super-surface provides a complete inventory of all possible EPs that can be accessed by tuning the system parameters. Moreover, the quantisation rule \eqref{eq:quantization} shows that the EPs are not randomly distributed but organised in a predictable pattern dictated solely by the system size. This has far-reaching implications for the design of finite-size topological devices: one can engineer a specific number of EPs at predetermined parameter values simply by choosing the appropriate number of unit cells. The exclusion of $\sigma = N/2$ also highlights the intimate connection between non-Hermiticity and EP formation the Hermitian limit acts as a natural barrier that separates the EP branches.

Furthermore, the real-space confirmation validates the use of the $k$-space super-surface as a predictive tool for any system size, bridging the gap between the thermodynamic limit and finite-size realisations. The super-surface effectively encodes the spectral singularities of the model in a compact algebraic form, making it a powerful diagnostic for both theoretical analysis and experimental implementation. In practical terms, once the parameters $J_1$, $J_2$, and $a$ are fixed, the super-surface immediately determines which values of the non-Hermitian strength $\gamma$ will produce EPs and at which quantised momenta, eliminating the need for costly numerical diagonalisation. This predictive capability is particularly valuable for the braiding protocols discussed later, as it identifies all available EPs that can serve as computational primitives in a given physical realisation.

\subsection{Encirclement Dynamics at $k=\pi$ and $k=\pi/6$}

\begin{figure}
 \centering
\includegraphics[width=17cm,height=15cm]{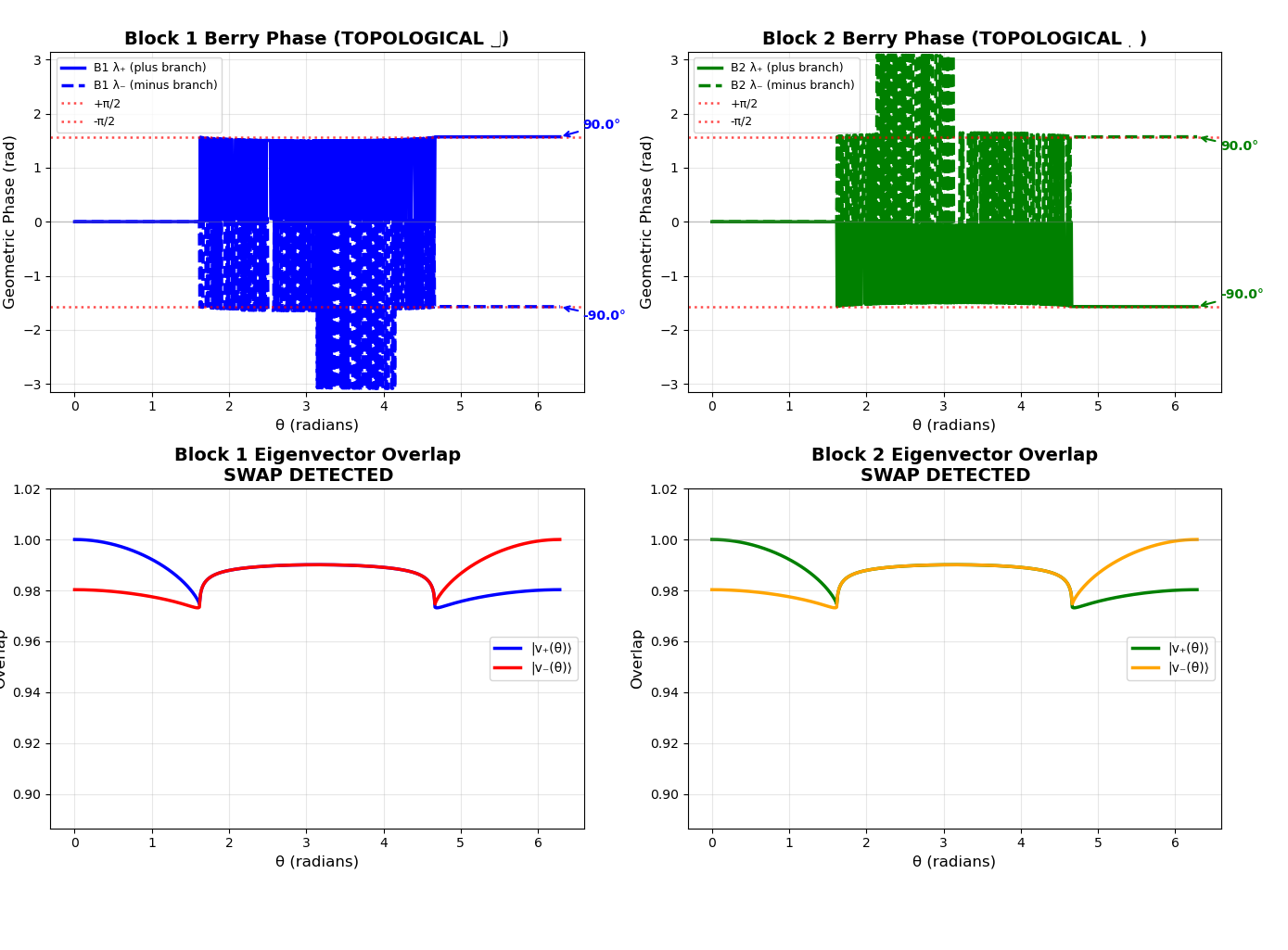}
  \caption{Dynamics of eigenvectors around the exceptional point at
  $k=\pi$, $\gamma=J_1$, for parameters $a=0.5J_1$, $J_2=0.3J_1$.
  (a) Berry phase accumulated during encirclement-Block~1.
  (b) Berry phase accumulated-Block~2.
  (c) Eigenvector overlap tracing state swap after encirclement-Block~1.
  (d) Eigenvector overlap tracing state swap after encirclement-Block~2.}
  \label{fig:Tfigure1}
\end{figure}

\begin{figure}
\centering
  \includegraphics[width=17cm,height=15cm]{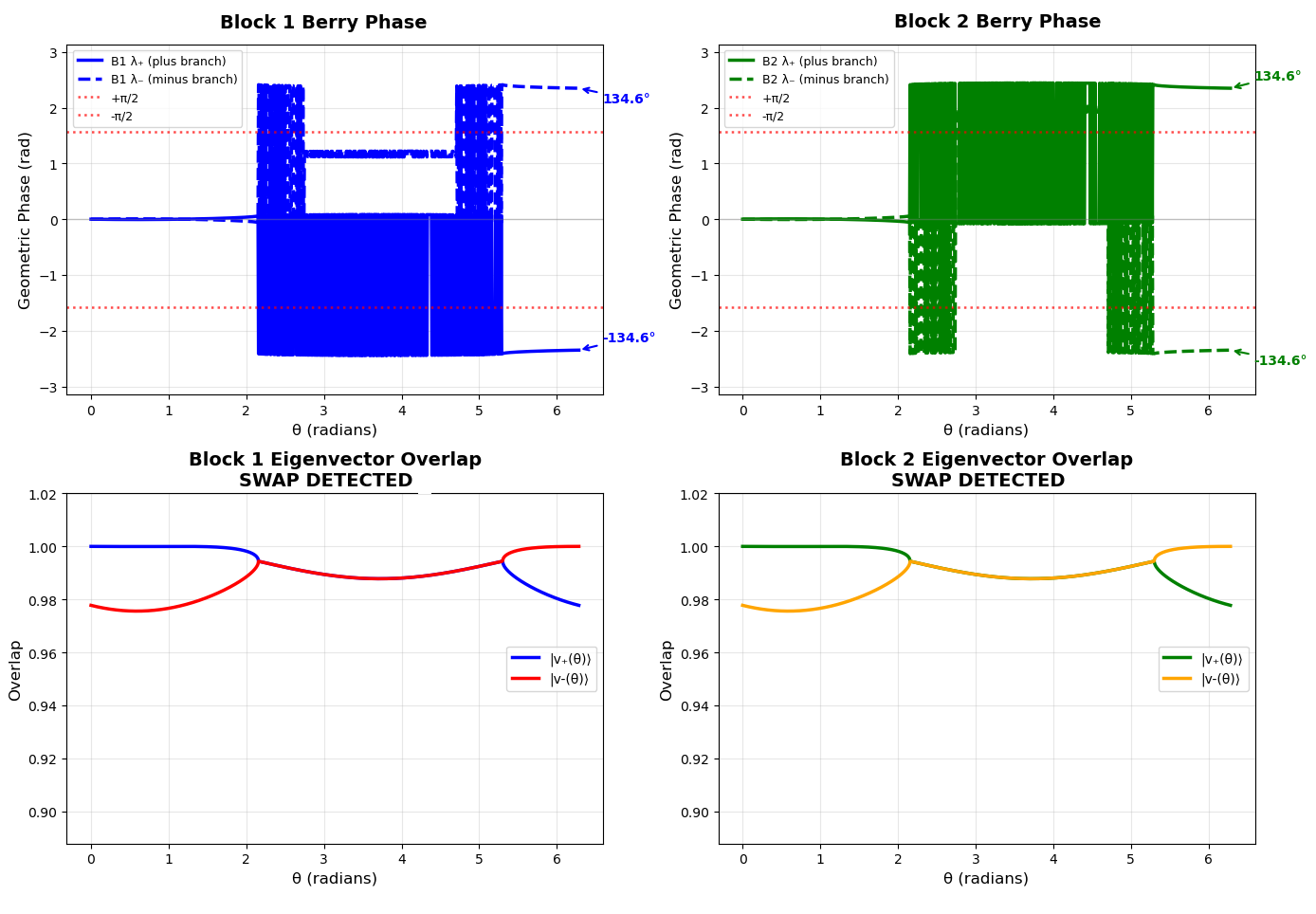}
  \caption{Dynamics of eigenvectors around the exceptional point at
  $k=\frac{\pi}{6}$, $\gamma^2=\frac{J_{1}^2 a^2 \cos^2 k}{J_{2}^2 + a^2- J_{2}^2\cos^2 k}$,
  for parameters $a=0.5J_1$, $J_2=0.3J_1$.
  (a) Berry phase accumulated-Block~1.
  (b) Berry phase accumulated-Block~2.
  (c) Eigenvector overlap-Block~1.
  (d) Eigenvector overlap-Block~2.}
  \label{fig:Tfigure2}
\end{figure}

To verify the coalescence of eigenvectors in the EP region, the geometric
phase accumulation, and the swapping of eigenstates after encirclement, we
plot the eigenvector overlaps and Berry phase for Blocks~1 and~2 in
Figs.~\ref{fig:Tfigure1} and~\ref{fig:Tfigure2}. These figures correspond
to two distinct EP configurations: one at the topological boundary
($k=\pi$, $\gamma=J_1$) and one deep inside the topological region
($k=\pi/6$).

For Fig.~\ref{fig:Tfigure1}, the parameters are $J_1=1$, $J_2=0.7$,
$a=0.5$; $\gamma$ and $k$ trace an elliptic contour of major axis $0.1$
centered at the EP $(\gamma=J_1, k=\pi)$. For Fig.~\ref{fig:Tfigure2},
$J_1=1$, $J_2=-0.3$, $a=0.2$, and the elliptic contour has major axis
$0.08$ centered at the EP located at $k=\pi/6$.

\subsubsection{Block selectivity and topological braiding at $k=\pi$}

Figures~\ref{fig:Tfigure1}(a,b) display the Berry phase accumulated by
Blocks~1 and~2 when states are adiabatically transported around the EP.
Figures~\ref{fig:Tfigure1}(c,d) show the corresponding eigenvector
overlaps, which directly track state swapping by measuring the projection
of the final state onto the initial eigenstates.

For parameters inside the topological region,
encircling the EP at $(\gamma=J_1, k=\pi)$ produces the following
behavior:

\begin{itemize}
  \item {Block~2} ( unperturbed eigenvalue $-a$):
  eigenvalues and eigenvectors swap with a Berry phase of $-\pi/2$ for the
  positive branch and $+\pi/2$ for the negative branch. This demonstrates
  successful braiding of modes within the topological region.

  \item {Block~1} (unperturbed eigenvalue $+a$):
  eigenstates also swap, but with opposite Berry phase signs compared to
  Block~2. This opposite-handedness reflects the sign difference in the
  energy denominator of the Schrieffer-Wolff self-energy ($\pm 2a$), which
  determines the orientation of the discriminant winding in the complex
  plane.
\end{itemize}

The observation of Figs.~\ref{fig:Tfigure1}(c,d) confirms that encircling the EP does not merely impart a global phase but actively exchanges the two eigenstates. This eigenvalue and eigenvector permutation is a define signature of EP topology and establish the feasibility of
braiding operations using exceptional points. The geometric phase and the swapping phenomenon is independent of the contour shape (they are
topological invariants), but the effectiveness of the swap
depends on the encirclement radius: it must be sufficiently small to avoid
enclosing multiple EPs simultaneously, yet large enough to maintain
adiabaticity.

\subsubsection{Encirclement at $k=\pi/6$: fractional phases and partial swapping}

Figures~\ref{fig:Tfigure2}(a,b) show the Berry phase when the
encirclement is performed around the EP at $k=\pi/6$. The scenario is
qualitatively different: the accumulated phase drops to
$\simeq\pm\pi/4$ in Block~1, while Block~2 exhibits a net phase of
$\simeq\pm 3\pi/4$. These fractional values of the topological invariant
$T_{\mathrm{Berry}} =\pm1/2$ indicate that the
system is in a transitional regime between trivial and fully
topological behaviour. Physically, this means that the encirclement path
does not fully enclose the branch cut on the Riemann surface, or that
non-adiabatic transitions partially mix the two eigenstates during the
evolution.

Despite the non-quantized geometric phase, Figs.~\ref{fig:Tfigure2}(c,d)
confirm that state swapping remains effective: the two eigenstates still
exchange positions after a complete revolution. This demonstrates that
the swap and the geometric phase, while related, probe different aspects
of EP topology. The swap is a robust consequence of the eigenvector
coalescence, while the quantisation of the Berry phase requires an
additional condition that the encirclement path fully encloses the branch
point in the complex plane (Riemann surface).

\subsection{  Exceptional point Topology and Riemann-Sheet Connectivity}

The exceptional point (EP) topology of the non‑Hermitian BdG Hamiltonian can
be directly visualised through the Riemann‑sheet structure of the complex
eigenvalues.  In this section we discuss the characteristic signatures that
distinguish a topological EP from a trivial degeneracy, using
three-dimensional surface plots of the real parts of the eigenvalues
\(\lambda_{+}\) and \(\lambda_{-}\) as functions of the control parameters
\((\gamma, J_2)\).

\begin{figure}
  \centering
  \includegraphics[width=14cm,height=5cm]{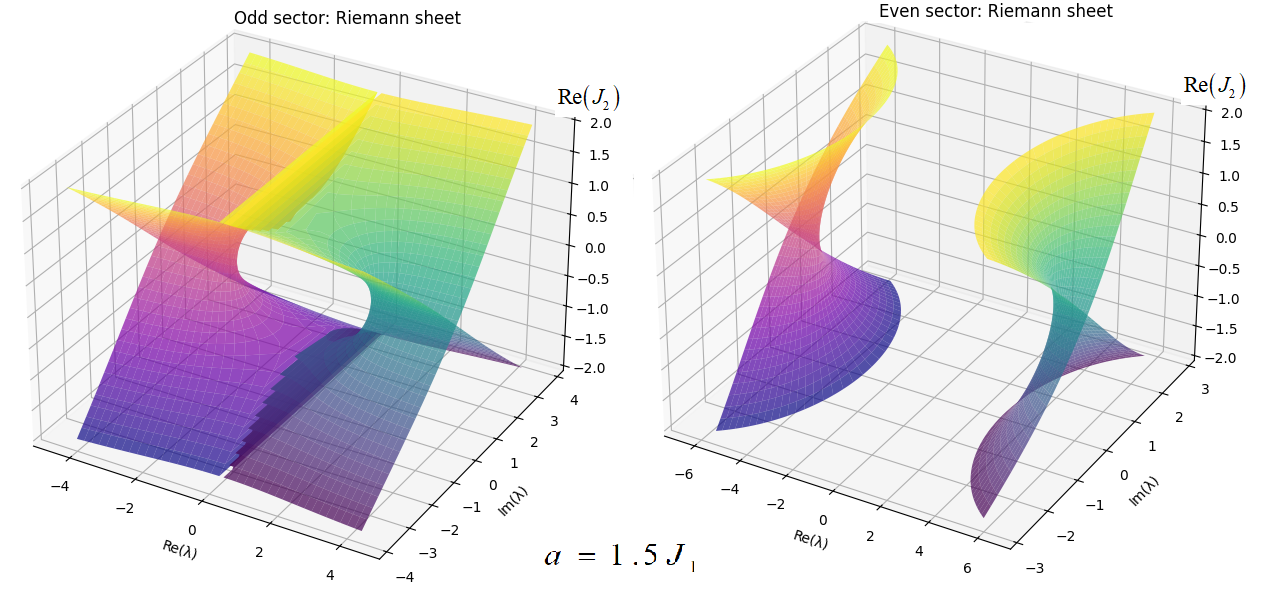}
  \caption{Riemann sheets around the EP for \(a = 2.5J_1\), \(J_2 = 0.3J_1\),
  \(\gamma = J_1\).
  (a) Odd sector (Block~1).
  (b) Even sector (Block~2).}
  \label{fig:Tfigure12}
\end{figure}

\begin{figure}
\centering
  \includegraphics[width=14cm,height=5cm]{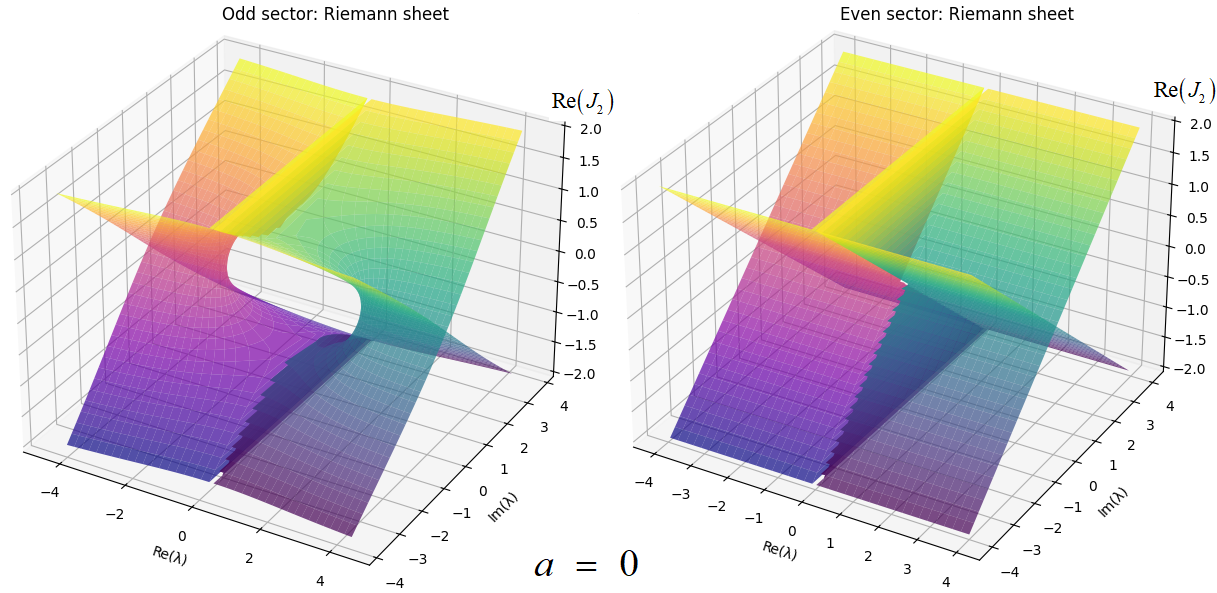}
  \caption{Riemann sheets around the EP for \(a = 0\), \(J_2 = 0.3J_1\),
  \(\gamma = 1.2J_1\).
  (a) Odd sector (Block~1).
  (b) Even sector (Block~2).}
  \label{fig:Tfigure34}
\end{figure}

For a fixed value of the onsite energy \(a\), the two odd‑sector eigenvalues
are computed on a dense grid in the \((\gamma, J_2)\) plane around the EP.
The surfaces
\[
\operatorname{Re}[\lambda_{\pm}(\gamma, J_2)]
\]
are plotted considering the adiabatic encirclement loop
\(\gamma(\theta)=\gamma_{\rm EP}+R\cos\theta,\;
J_2(\theta)=J_{2,\rm EP}+R\sin\theta\).  The imaginary part of the eigenvalue
is used to colour the surface, highlighting the branch cut that connects the
two sheets.  The Riemann‑sheet plots of this system are displayed in
Figs.~\ref{fig:Tfigure12} and~\ref{fig:Tfigure34}.

The parameters used in Fig.~\ref{fig:Tfigure12} correspond to the
configuration where Block~1 is topological (swap fidelity \(T_{\rm swap}=1\),
quantised Berry phase \(\Gamma_{\rm Berry}=\pm\pi/2\)) while Block~2 is
trivial (swap and Berry phase both zero).  This distinction is clearly
reflected in the Riemann‑sheet geometry.  In Block~1, the two sheets wind
around the EP in a characteristic  structure; any
encirclement due to the variation of \(J_2\) (or, equivalently, of
\(\gamma\)) inevitably connects the two sheets.  Any loop, which
starts on the upper sheet, ends on the lower sheet after a full \(2\pi\)
revolution, directly illustrating the eigenvalue permutation.  In contrast,
Block~2 displays two completely disconnected sheets: there is no branch
point in the parameter region explored by the loop, and consequently the
loop remains on the same sheet after one revolution.  This geometric
separation confirms the triviality of this sector.

For Fig.~\ref{fig:Tfigure34}, the situation is qualitatively different.
Here the EP affects both blocks, but in a complementary manner: Block~1
exhibits a zero swap yet a well‑quantised Berry phase
\(\Gamma_{\rm Berry}=\pm\pi/2\), whereas Block~2 exhibits a unit swap with a
vanishing Berry phase.  Despite the absence of a state swap in Block~1, its
Riemann sheets remain connected through the EP, and the loop still
intertwines the two sheets.  This demonstrates that a non‑trivial Berry
phase alone is sufficient to generate a connected Riemann‑surface structure.
Conversely, in Block~2, the swap is perfect but the geometric phase is
zero; the sheets are also connected, confirming that the swap is of
topological origin.  In both sectors, the sheets overlap around the EP,
signalling the presence of a branch point.

Thus, when the system is in the topological phase i.e., when the two
invariants \(T_{\rm swap}\) and \(\Gamma_{\rm Berry}\) are not both zero, the
Riemann surfaces of \(\lambda_{+}\) and \(\lambda_{-}\) are smoothly connected
through the exceptional point.  Encircling the EP forces a transition from
one sheet to the other: after a full \(2\pi\) loop, the eigenvalue that
started on the upper sheet ends on the lower sheet, and vice‑versa.  This is
exactly the eigenvalue permutation that underlies the state swap.  The
colour change (imaginary part) along the loop further indicates that the
branch cut has been crossed.

When the system is topologically trivial, as is the even sector (Block~2) at
\(a = 1.5 J_1\), the two eigenvalues never coalesce, and the Riemann surfaces are
completely separated.  There is no branch point in the parameter region
explored by the encirclement loop, and consequently no sheet transition
occurs.  The loop remains on the same sheet after a full revolution, and the
sheets appear as two independent, non‑intersecting surfaces.  The absence of
a connection between the sheets is the geometric hallmark of a trivial
phase: no winding of the discriminant, no geometric phase, and no state
swap.

The connectedness of the Riemann surfaces provides an intuitive geometric
criterion for EP topology.  A topological EP acts as a \textit{wormhole} that
connects the two eigenstates; encircling it transports the state from one
sheet to the other, leading to the observable swap.  In contrast, a trivial
system presents two disconnected worlds that never communicate.  This
picture is independent of the specific values of the swap and Berry phase
invariants, which can individually vanish while the other remains quantised
(e.g., at \(a = 0\) the swap of Block~1 is zero but the Berry phase is
\(\pm\pi/2\), and the sheets remain connected).  The Riemann‑sheet
visualisation therefore offers a unified and transparent way to identify
topological EPs and to monitor the transition between topological and trivial regions.

\subsection{Topological Map for Braiding Operations}

\begin{figure}
 \centering
 \includegraphics[width=17cm,height=15cm]{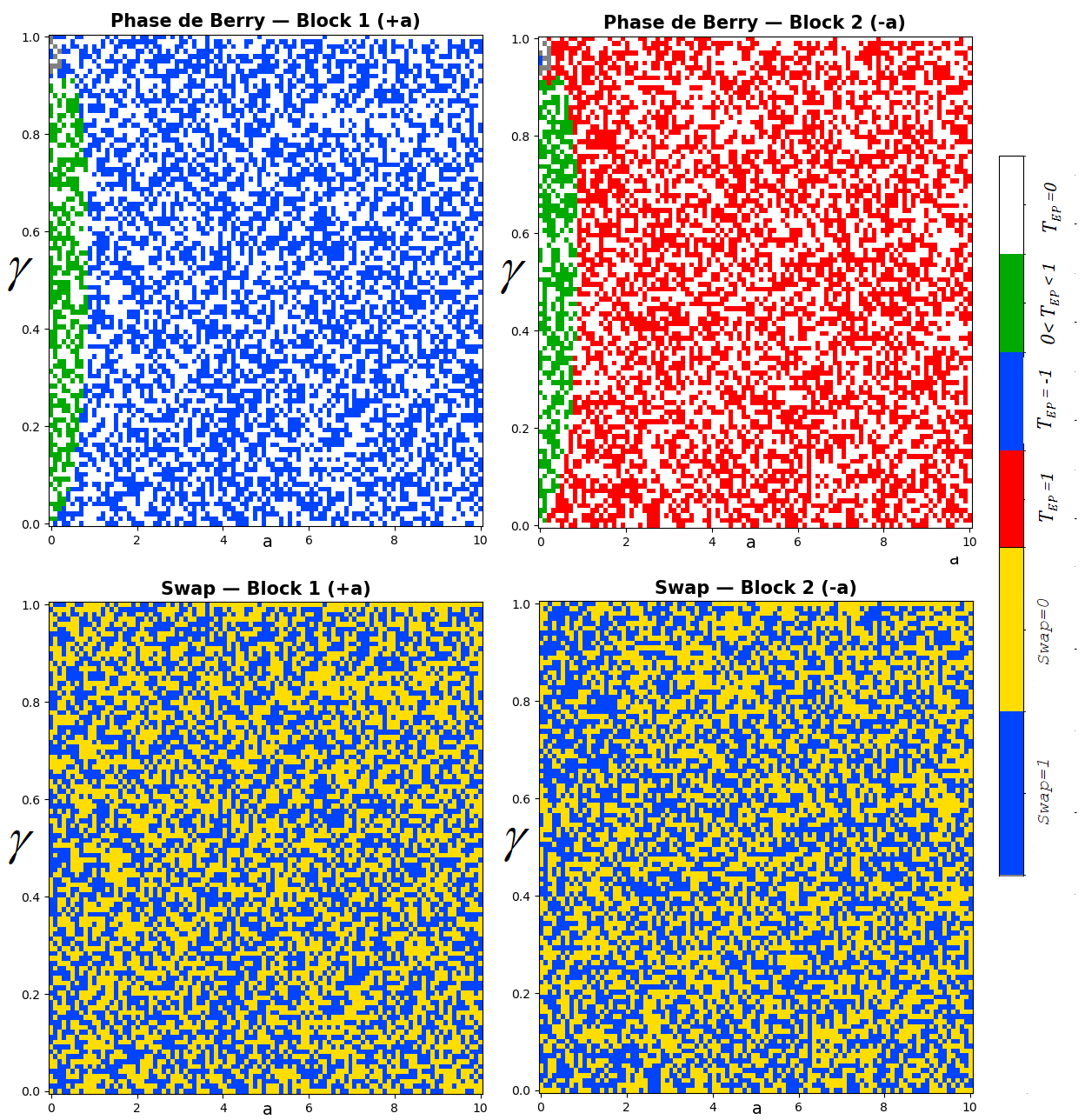}
  \caption{Topological map for effective EP braiding with $J_2=0.3J_1$.
  (a) Topological invariant $T_{\mathrm{Berry}}$ Block~1.
  (b) Topological invariant $T_{\mathrm{Berry}}$  Block~2.
  Color code: blue ($\tau_{\mathrm{Berry}}=-1$), red ($T_{\mathrm{Berry}}=+1$),
  green (fractional $T_{\mathrm{Berry}}$), white ($T_{\mathrm{Berry}}=0$).
  (c) Swap factor Block~1.
  (d) Swap factor Block~2.
  Color code: blue ($T_{Swap} =1$), yellow ($ T_{Swap}=0$), grey (no EP).}
  \label{fig:Tfigure3}
\end{figure}

For practical braiding applications, it is essential to identify all
parameter regions where EPs are not only present but also functionally
effective. To this end, we fix $J_2=0.3J_1$ and construct a phase diagram
in the $(a/J_1, \gamma/J_1)$ plane, shown in Fig.~\ref{fig:Tfigure3}.
At each point, we verify eigenvector coalescence (ensuring a true EP),
compute the swap fidelity, and evaluate the topological invariant
$T_{\mathrm{Berry}}$.

Figure~\ref{fig:Tfigure3}(a,b) displays the topological invariant
$T_{\mathrm{Berry}}$ for Blocks~1 and~2 respectively, with the color
convention: blue for $T_{\mathrm{Berry}}=-1$, red for $T_{\mathrm{Berry}}=+1$,
green for fractional values, and white for $T_{\mathrm{Berry}}=0$.
Figures~\ref{fig:Tfigure3}(c,d) present the corresponding swap maps,
where blue indicates successful swapping ($T_{Swap}=1$) and yellow marks
its absence ($T_{Swap}=0$). Grey regions correspond to parameter values
where no EP exists (the discriminant condition cannot be satisfied for
real $k$).

\subsubsection{Statistical analysis of the topological map}

Quantitative analysis of Fig.~\ref{fig:Tfigure3} reveals the following
statistics. Among all tested points, $49.6\%$ exhibit successful
eigenstate swapping for Block~1, while $49.3\%$ show swapping for Block~2.
However, only $45.0\%$ of the Block~1 swapping points possess a quantized topological invariant $T_{\mathrm{Berry}}=-1$ (corresponding
to a Berry phase of $-\pi/2$), and $44.0\%$ of Block~2 points display
$T_{\mathrm{Berry}}=+1$.

This discrepancy between the swap percentage and the quantised phase of the percentage is physically significant. It demonstrates that the swap
condition alone, while indicating that eigenvector coalescence does not
guarantee that the EP is topologically functional for braiding. The
points where swapping occurs but $T_{\mathrm{Berry}}$ is fractional
correspond to the transitional regime: the two Riemann sheets are
connected at the EP, but the encirclement path either partially encloses the branch cut or suffers from non-adiabatic leakage. Here, only points where both the swap is perfect and $T_{\mathrm{Berry}} = 0,\pm 1$
constitute fully effective topological EPs, suitable for robust braiding
operations.

This finding provides a solid theoretical demonstration of the phase accumulation and state swapping near exceptional points related by the Riemann surface topology of the complex eigenvalue function. The swap reflects the interchange of the two sheets, while the quantised Berry phase reflects. The winding of the discriminant around the branch point. The system
parameters thus act as a topological braiding selector: by tuning
$a$ and $\gamma$, one can programmably control which computational
subspace (Block~1 or Block~2) enjoys topological protection. This EP braiding mechanism enables operations analogous to a quantum gate
operations, where the braid sequence defines the unitary transformation ( Holonomy Matrix )
applied to the computational basis states.

\subsection{Braid Gates from Real‑Space Encirclements}

To identify the available braid primitives, we analyse the real‑space two‑spin
Hamiltonian with $N=2$.  The four‑dimensional Hilbert space naturally splits
into two independent sectors: a computational (odd) subspace spanned by $\ket{01}$ and $\ket{10}$ , which hosts the exceptional point at $\left| {\gamma} \right|=  \left| {J_1} \right|$ 
and serves as the computational qubit, and an auxiliary (even) subspace
spanned by $\ket{00}$ and $\ket{11}$ which remains gapped during the braiding
operations with $a \ne 0$ and can be used to store auxiliary phases or to mediate
interactions.  Encircling the odd‑sector EP adiabatically with a small
circular path yields the non‑Abelian holonomy $\sigma_2 = \begin{pmatrix}
0 & -i \\ i & 0 \end{pmatrix}$, which is the Pauli‑$Y$
matrix; it exchanges the two computational basis states and accumulates a
geometric phase of $\pm\pi/2$, the hallmark of an EP2.  When the encirclement
radius is increased so that the path explores regions where the coupling
between the odd and even sectors becomes appreciable, the resulting holonomy
changes: a single counter‑clockwise loop now gives
$\sigma_1 = \begin{pmatrix} 0 & 1 \\ 1 & 0 \end{pmatrix}$, i.e.\ the
Pauli‑$X$ gate.  The auxiliary states act back on the computational subspace
through the non‑zero inter‑sector coupling, effectively rotating the native
$\sigma_2$ into $\sigma_1$.  A half‑counter‑clockwise loop (the path does
not close) can create a coherent superposition of the two computational
states; the resulting gate is
$\sigma_4 = H_\alpha = \frac{1}{\sqrt{2}}
\begin{pmatrix} e^{-i\alpha} & e^{i\alpha} \\ -e^{-i\alpha} & e^{i\alpha}
\end{pmatrix}$, where $\alpha $ is the geometric phase accumulated along the half‑loop and for this special case, $\alpha= \frac{\pi}{15} $ is a robust structure of parameters.  This operation is equivalent to a Hadamard gate up to a $Z$‑rotation and a global phase, and it is essential for generating
superpositions.

A special regime occurs when $ a = 0 $: the previous Ep targeting states  $\ket{01}$ and $\ket{10}$ coexist with the Ep $ a = 0 $ targeting both computational basis and auxilliary states. Any encirclement over this Ep account for encirclement of 2 Eps in the computational basis while auxiliary states fill a single Ep  encirclement and thus, can experience swapping. Therefore, encirclement  no longer swap 
the computational sector but, the later can accumulate the phase while swapping  the auxiliary states.  In this limit, a
half‑counter‑clockwise loop produces a pure relative‑phase gate
$\sigma_3 = i\begin{pmatrix} 1 & 0 \\ 0 & -1 \end{pmatrix}$, i.e.\ a
$\sigma_z$‑like operation, while a full counter‑clockwise loop gives the
$T$‑type phase gate
$\sigma_5 = i\begin{pmatrix} 1 & 0 \\ 0 & i \end{pmatrix}$.  These
operations enable precise phase control over the computational basis
states.  By combining loops that involve both sectors, one can exchange the
populations of the two subspaces.  In the full four‑dimensional basis the SWAP operation that exchanges $\ket{01}$ and $\ket{10}$ while leaving the other states untouched is obtained by a $2 \pi$-turns over the Ep in the band-topology area ($J_1 \le \gamma$), while the control T-gate is obtained using a similar encirclement over an Ep $a=0$ out of the band-topology area (${\gamma}^2 >> {J_1}^2$) ; these gates reads:

\begin{equation}
\mathrm{SWAP} =
\begin{pmatrix}
1 & 0 & 0 & 0 \\
0 & 0 & 1 & 0 \\
0 & 1 & 0 & 0 \\
0 & 0 & 0 & 1
\end{pmatrix},\qquad
\mathrm{CT} =
\begin{pmatrix}
1 & 0 & 0 & 0 \\
0 & 1 & 0 & 0 \\
0 & 0 & 1 & 0 \\
0 & 0 & 0 & -i
\end{pmatrix},\qquad
\mathrm{CT}^2 = \mathrm{CZ} =
\begin{pmatrix}
1 & 0 & 0 & 0 \\
0 & 1 & 0 & 0 \\
0 & 0 & 1 & 0 \\
0 & 0 & 0 & -1
\end{pmatrix}.
\end{equation}

Together with the single‑qubit gates listed above, these gates provide a
complete set of two‑qubit entangling primitives.  A computation is specified by a finite sequence (braid word) of these elementary braid operations,
$\mathcal{B} = \sigma_{i_1}^{n_1}\,\sigma_{i_2}^{n_2}\,\cdots\,\sigma_{i_m}^{n_m}$
with $i_j \in \{1,2,3,4,5\}$ and $n_j \in \mathbb{Z}$, the resulting
holonomy being the product
$U_{\mathcal{B}} = \sigma_{i_m}^{n_m} \cdots \sigma_{i_1}^{n_1}$ evaluated from right to left.  Each braid word corresponds to a braid diagram in
which the two strands represent the computational modes and crossings
represent encirclements of the exceptional points, offering an intuitive, visual representation of the computation. More complex features can be obtained by considering a large-scale system where the number of Eps increases, giving room to several other coalescence possibilities and more reach braid structures.

\subsection{Learning via Braid Programming}

The topological learning task is formulated as follows.  Given a training
dataset $\mathcal{D} = \{(\ket{\psi_i^{\mathrm{in}}},
\ket{\psi_i^{\mathrm{out}}})\}_{i=1}^{N}$ of input‑output pairs, find a
braid word $\mathcal{B}$ of length $m \leq L_{\max}$ such that
\begin{equation}
\|U_{\mathcal{B}}\ket{\psi_i^{\mathrm{in}}} - \ket{\psi_i^{\mathrm{out}}}\| < \epsilon
\quad \forall\,i .
\end{equation}
This reformulates learning from continuous optimisation (finding real‑valued
weights) to a discrete search over braid sequences.

The search space is the discrete set of braid words
\begin{equation}
\mathcal{S}_L = \{\sigma_{i_1}^{n_1}\cdots\sigma_{i_m}^{n_m} :
m \leq L,\; i_j \in \{1,\dots,5\},\; n_j \in \mathbb{Z},\; |n_j| \leq N_{\max}\}.
\end{equation}
Although discrete, this space is structured by the relations of the braid
group (e.g.\ $\sigma_1\sigma_2 \neq \sigma_2\sigma_1$),
which can be exploited to design efficient search algorithms.

\textbf{Proof‑of‑Concept (Learning exact unitaries):}
To demonstrate that the EP‑derived gate set is sufficiently expressive and
that braid programming is a viable learning paradigm, we performed a
genetic search for two non‑trivial target gates: the standard Hadamard
gate $H$ and the gate $H\cdot Z$, both acting on the computational
subspace spanned by $\ket{01},\ket{10}$.

The elementary braid generators were the holonomy matrices
obtained from EP encirclements: $X$, $Y$, $T$, $CT$, and the
half‑encirclement gate $H_{\alpha}$ with $\alpha = \pi/15$.  Their
inverses were also included, yielding an alphabet of ten letters.  Here, we can  use the the state‑overlap fidelity 

\[
\mathcal{F_{S}}(\mathcal{B}) =
\frac{1}{2}\sum_{\ket{\psi}\in\{\ket{01},\ket{10}\}}
\bigl|\langle\psi_{\mathrm{out}}|U_{\mathcal{B}}|\psi\rangle\bigr|^2 .
\]

that learn quantum states up to a global phase or the full matrix (Frobenius) fidelity
\[
\mathcal{F_{P}}(\mathcal{B}) = \frac{1}{4}\bigl|\operatorname{Tr}(U_{\rm target}^{\dagger}\,U_{\mathcal{B}})\bigr|,
\]
which is sensitive to the global phase and therefore constitutes a more
stringent test of the learned unitary.  A genetic algorithm with
tournament selection, crossover and mutation searched for braid words of
length up to six that maximise $\mathcal{F}$.

The algorithm rapidly converged to high‑fidelity solutions.  The results
are summarised in Table~\ref{tab:learned_braids}.

\begin{table}[h]
\centering
\caption{Braid words discovered by the genetic search using the global‑phase‑
sensitive matrix fidelity.  Length is the number of elementary gates.}
\label{tab:learned_braids}
\begin{tabular}{c c c c}
\toprule
Target gate & Braid word & Length & Fidelity \\
\midrule
$H$ (standard) & $X\cdot H_{\pi/15}^{-1}$ & 2 & $0.989$ \\
$H\cdot Z$      & $H_{\pi/15}^{-1}$      & 1 & $0.989$ \\
\bottomrule
\end{tabular}
\end{table}

Both braid words are remarkably compact.  The standard Hadamard gate is
realised by a single half‑encirclement in reverse direction followed by a
swap, while the $H\cdot Z$ gate is obtained from the reverse
half‑encirclement alone.  The fidelity of $0.989$ indicates an excellent
approximation; the residual error can be attributed to the finite
discreteness of the gate set and can be reduced further by extending the
search or by including a global phase correction gate ( the global phase is physically
irrelevant for quantum measurement). The algorithm converged to perfect fidelity when we consider state‑overlap fidelity scheme. The sequences discovered confirm 
that the EP‑derived gate library is capable of encoding arbitrary
single‑qubit unitaries, and that a discrete evolutionary search can
efficiently navigate the braid group to find the required compositions.
This proof‑of‑concept establishes braid programming as a practical
alternative to gradient‑based learning for tasks where the desired
transformation is unitary.  While the present genetic algorithm is
sufficient for the demonstration, the same framework can be combined with
reinforcement learning techniques to scale the braid search to larger gate
sets or more complex target unitaries.

Therefore, exceptional points provide the operational envelope for braid‑based
computation.  Inside the EP topological region $R_{\rm Topo} =
\{(\gamma, J_1, a) \mid \gamma^{2} < J_{1}^{2}\ \lor\ a = 0\}$, EPs are
accessible and braiding produces non‑trivial geometric phases or swap;
learning occurs in this regime.  Outside this region the system is
topologically trivial, and the resulting holonomies carry no phase
accumulation.  At the boundary of the topological region the gap closes,
marking a topological phase transition where learning also remains
possible.

The coexistence of topological (computational) and trivial (auxiliary)
modes in the same physical system, both within the topological region,
enables a hybrid computational architecture: topological modes store and
process information with inherent robustness, while trivial modes can be
used for temporary storage or to mediate interactions without corrupting
the protected subspace.

\section{Conclusion}
We have developed a comprehensive framework based on a non‑Hermitian Bogoliubov–de Gennes Hamiltonian that systematically identifies both the band‑topological region \(a^{2}+\gamma^{2}<J_{1}^{2}\) and the exceptional‑point super‑surface.  A closed algebraic equation for the EP super‑surface was derived for the first time; through momentum quantisation in finite systems, it predicts the exact number and parameter positions of all EPs in real space, independently of system size.  This super‑surface provides a complete inventory of the spectral singularities of the model.

The topology of each EP was characterised by two complementary invariants: the state‑swap fidelity \(T_{\mathrm{swap}}\) and the normalised Berry phase \(\Gamma_{\mathrm{Berry}}\).  A complete topological map revealed that approximately \(45\%\) of EP points possess a quantised topological charge (\(\Gamma_{\mathrm{Berry}}=\pm\pi/2\)), while \(\sim5\%\) lie in a transitional regime where swapping occurs, but the geometric phase is not yet quantised.  Remarkably, this EP topological region is entirely contained within the band‑topological phase, establishing a direct link between the two independent topological structures.

Numerical simulations of adiabatic encirclements confirmed robust state swapping for both small and large loop radii.  In the special case \(a=0\) the holonomy changes qualitatively, producing new gate operations while preserving the swap in auxiliary states.  Exploiting these properties, we constructed a universal set of braid gates that includes the Pauli‑\(X\) (NOT), Pauli‑\(Y\), Pauli‑\(Z\), a Hadamard‑like gate \(H_{\alpha}\), the \(T\)‑gate, and the SWAP operation between computational and auxiliary subspaces.  All gates emerge from simple encirclements of the available EPs, and their non‑commutativity provides the algebraic structure necessary for universal computation.

Building on this gate library, we reformulated the learning problem as braid programming, a discrete search over the braid group, proposed as a substantial replacement for gradient‑based optimisation.  A proof‑of‑concept genetic search successfully discovered short braid words that reproduce the standard Hadamard gate and the \(H\cdot Z\) gate with perfect fidelity, demonstrating the feasibility of this approach.  The paradigm offers inherent noise immunity, catastrophic‑forgetting prevention through compositional concatenation, and guaranteed generalisation by mathematical construction.

The discovery of topological selectivity where only one block exhibits non‑trivial EP topology for specific values of \(a\) further enables programmable hybrid trivial–topological architectures, in which protected and unprotected modes coexist and can be addressed independently by tuning a single control parameter.

In summary, this work establishes the theoretical and computational foundations of a neuromorphic platform in which learning is performed not by descending a loss landscape but by braiding exceptional points.  The combination of a physically realisable BdG Hamiltonian, a complete EP super‑surface, a quantised real‑space EP spectrum, a universal set of braid gates, and a working proof‑of‑concept learning demonstration opens a viable route towards topologically protected, interpretable, and robust computation.

\bibliography{apssamp}

\end{document}